# Giant Resonance Raman Scattering via Anisotropic Excitons in ReS$_2$

Pritam Das, Devarshi Chakrabarty, Neha Gill, Sajal Dhara*

Department of Physics, Indian Institute of Technology, Kharagpur, India

**Abstract:**

Anisotropic two-dimensional (2D) semiconductors recently have emerged as a promising platform for polarization-controlled Raman amplification. In this study, we probe energy-dependent resonant Raman scattering in few layer ReS$_2$ under different polarization configurations. We identify two distinct excitation regimes, each characterized by a resonance condition where either the pump or the Stokes photon energy aligns with an excitonic transition. A two-order-of-magnitude enhancement in Raman intensity is observed when the pump energy is tuned near the exciton resonance. Under Stokes-resonant conditions, additional Raman lines accompanied by excitonic photoluminescence are observed, suggesting the participation of non-Bloch intermediate states in the scattering process. These findings shed light into the influence of excitons in modulating nonlinear optical phenomena in anisotropic 2D materials, offering valuable insights for the design of tunable photonic and optoelectronic devices based on anisotropic layered materials.

Anisotropic two-dimensional (2D) materials enable polarization-dependent light–matter interactions and offer promising applications in polarization-controlled optical devices such as all optical switches[1], polarized LEDs[2], photodetectors[3], and quantum logic gates[4]. Among anisotropic 2D materials, rhenium disulfide ($ReS_2$) has attracted considerable attention due to its unique optical, vibrational, and electronic properties arising from its distorted 1T crystal structure. The reduced crystal symmetry of $ReS_2$ leads to polarization dependent multiple Raman modes and pronounced anisotropic excitons[5–12]. Understanding Raman scattering is important in van der Waals materials not only for its significance in characterization[13] such as layer thickness[14–16], phase transition[17], stacking order[15,18,19] and twist angle[20], but also Raman processes recently has been utilized to achieve zero threshold Raman lasing[21] in anisotropic exciton-polariton system with non-Hermitian topological band structure[22,23]. A rich spectrum with 18 Raman modes has been identified in earlier studies in the range of 100-450 cm$^{-1}$ in $ReS_2$[24–32], however, a comprehensive understanding of the Raman amplification in this anisotropic 2D layer remains unexplored.

In this work, the excitation energy was systematically tuned across the excitonic resonances, leading to a giant enhancement in Raman scattering (RS) efficiency by a factor of 100–200 under resonant conditions as compare to observed isotropic TMDs such as $MoS_2$[33,34], $MoSe_2$[35], and $WS_2$[36]. Moreover, the emergence of new Raman modes along with excitonic photoluminescence highlights the critical role of excitons in the RS process. A schematic of the backscattering geometry for a single-layer distorted 1T-phase[37] $ReS_2$, characterized by two nearly orthogonally polarized excitons and phonons, is shown in Fig. 1(a). The optical image of the exfoliated flake is shown in the inset of Fig. 1(a), with the thickness estimated to be approximately 8 nm based on atomic force microscopy measurements (see Fig. S1 in Supporting Information). The crystallographic orientation of the flake can be inferred from the sample edges, as $ReS_2$ typically cleaves along the covalent Re-S bonds, which are aligned with the crystallographic b-axis, indicated by the arrow in the inset of Fig. 1(a). The low symmetry and quasi-indirect bandgap[5,8,38] nature of $ReS_2$ makes it an excellent candidate for Raman scattering studies. Typically, Raman scattering efficiency is enhanced when virtual energy levels closely match real electronic energy levels, a phenomenon known as Resonance Raman Scattering[39](RRS). In two-dimensional materials like $ReS_2$, excitons play a significant role in the light-matter interaction leading to amplified Raman scattering efficiency when virtual energy states lie in the vicinity of the excitonic energy levels. Fig. 1(b) schematically illustrates

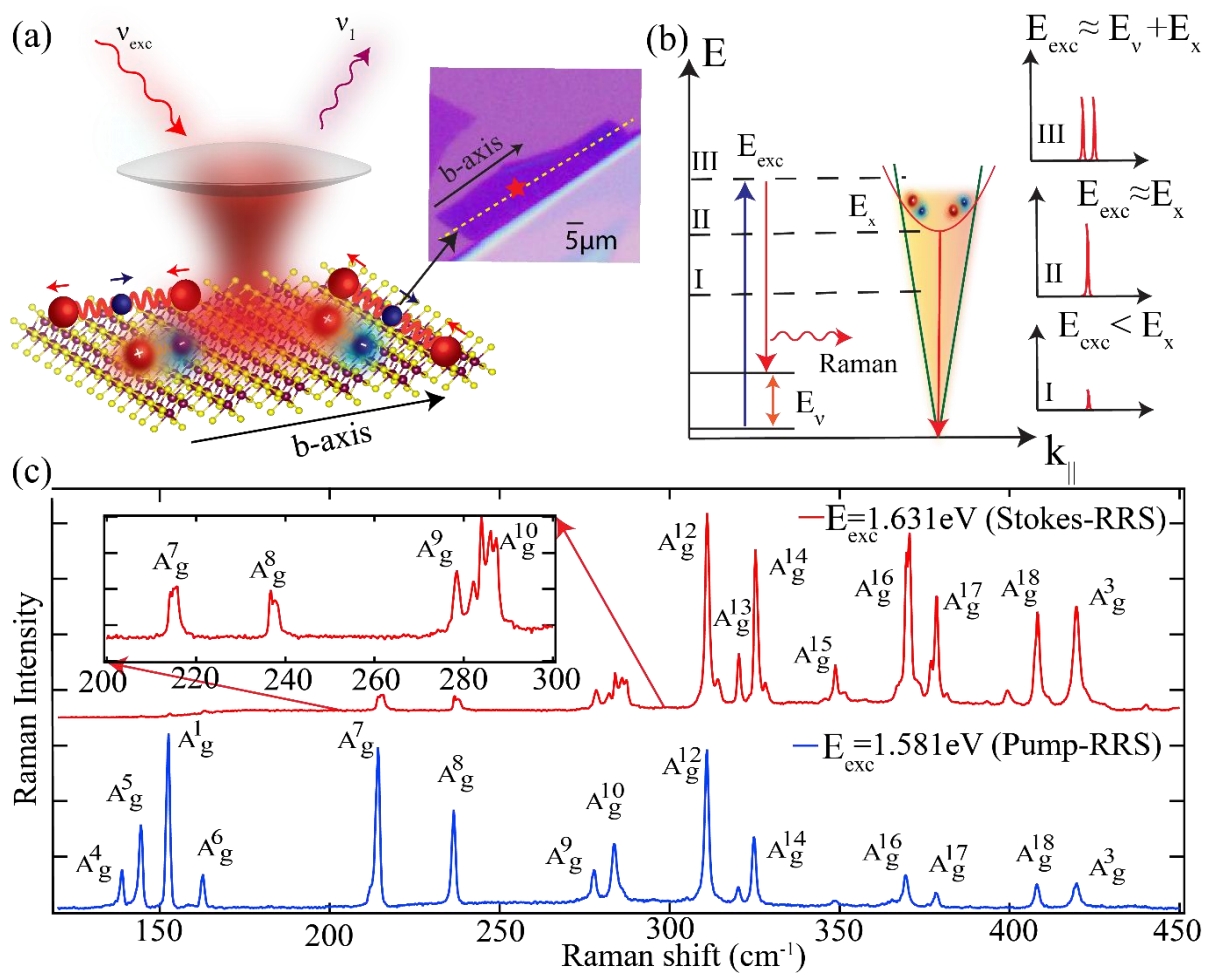

Fig.1 (a) Schematic illustration of a focused laser beam exciting few-layer $ReS_2$, highlighting linearly polarized excitons and phonons, with a zoomed view of the optical image as the inset.(b) An illustration of Raman scattering overlaid on excitonic dispersion, showing three representative excitation energies (I, II, III), marked by dotted lines, corresponding to non-resonant, pump-resonant, and Stokes-resonant Raman scattering conditions. The right panel presents a comparative Raman intensity plot and the emergence of new Raman lines across the three excitation regimes. (c) Raman spectra of 8 nm thick $ReS_2$ measured under pump and Stokes-resonant condition using excitation energies of 1.581eV (blue) and 1.631eV (red), respectively.

the influence of excitons on the Raman scattering process, depicting three virtual energy levels (I–III). When the pump energy lies below the exciton resonance energy ($E_X$) conventional spontaneous Raman scattering takes place. However, as the excitation energy approaches the exciton resonance ($E_{exc} = E_X$), a pronounced enhancement in Raman scattering is observed, referred to as pump-resonant Raman scattering (pump-RRS). Furthermore, when the excitation energy is increased such that the Stokes-shifted photon energy aligns with the exciton resonance ($E_{exc} = E_X + E_v$), an additional enhancement in Raman scattering efficiency is observed, accompanied by the appearance of new Raman modes, a phenomenon known as Stokes-resonant Raman scattering (Stoke-RRS). Polarization- dependent RRS is also observed, arising from the strong anisotropy of excitons, as discussed in detail in a later section. The Raman spectra corresponding to pump-RRS and Stokes-RRS excitations are observed at 1.581 eV and 1.631eV respectively, are shown in figure 1(c) for the input polarization oriented nearly along the b-axis. The reasoning behind why these specific excitations fulfil pump and stoke RRS condition is discussed in a subsequent section. For an excitation energy of $E_{exc} = $ 1.581 eV, 16 distinct Raman modes are observed, identified, and labelled according to previous studies. In contrast, at an excitation energy of $E_{exc} = $ 1.631 eV, additional Raman modes emerge alongside previously observed modes. A magnified spectral window in the range of 200–300 cm$^{-1}$ for the 1.631 eV excitation is shown in the inset of Fig. 1(c). Under stokes-resonant conditions, two closely spaced peaks (Δ~3 cm$^{-1}$) appear near the $A_g^7$, $A_g^8, A_g^{16}, and\ A_g^{17}$ modes, along with four closely spaced features near the $A_g^{10}$ mode. The spectral linewidths (~0.8cm$^{-1}$) of these Raman peaks are comparable to the resolution limit of the spectrometer. For a comprehensive understanding of excitonic effects on Raman scattering, the excitonic properties has been characterized by utilizing reflectivity and photoluminescence (PL) spectroscopy, as shown in Fig. 2(a) and 2(b). At room temperature, the excitonic features in both PL and reflectivity spectra are excessively broad, preventing the resolution of individual peaks. To address this limitation, the sample was cooled utilizing a closed-cycle cryostat, and all optical measurements were subsequently conducted at 3.2 K. Reflectance spectra were obtained using a broadband halogen light source with a spot size of approximately 3 μm, whereas significant photoluminescence was observed under excitation at 1.88 eV, a value considerably higher than the exciton resonance energies. In Figure 2(a), we present the differential reflectance, defined as, $\frac{\Delta R}{R} = \frac{R_{sample} - R_{sub}}{R_{sub}}$, where $R_{sample}$ and $R_{sub}$ are the reflectance spectra from the sample on the substrate and the bare $SiO_2$/Si substrate, respectively. We observed two prominent peaks at $1.545 \pm 0.001$ eV and $1.585 \pm 0.001$ eV

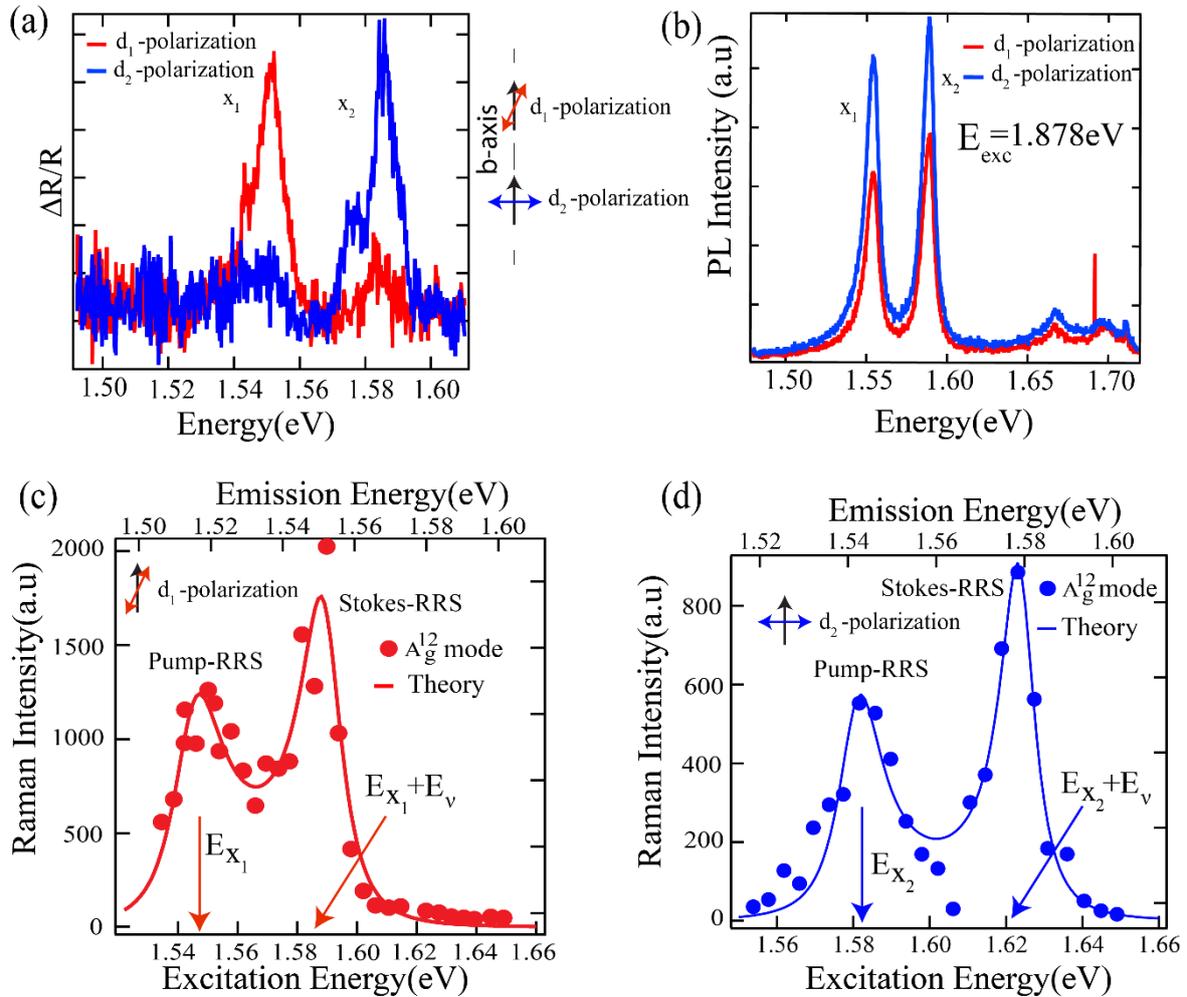

Fig.2 (a) Differential reflectance spectra for two incident polarization directions, $d_1$ (red) and $d_2$ (blue), defined with respect to the b-axis (as indicated between panels (a) and (b)), showing two excitonic resonances in 8 nm $ReS_2$. (b) Polarization-resolved photoluminescence intensity measured under $d_1$(red) and $d_2$(blue) polarized excitation. (c) Variation of the integrated Raman intensity of the $A^{12}g$ mode with the excitation energy for $d_1$ polarization. (d) Same as (c) but for $d_2$ polarization.

for the input polarization nearly parallel and perpendicular to the b-axis, respectively, which were identified as $X_1$ and $X_2$ excitons[5–7]. Polarization-resolved measurements reveal that the two excitonic species, $X_1$ and $X_2$, are polarized at angles of approximately 10° and 90°, respectively, with respect to the b-axis. These values are consistent with previously reported results within experimental error[5,6,38,40]. These two polarization are denoted as $d_1$ and $d_2$, respectively. Notably, unlike the reflectance spectra, both excitons were detected for both $d_1$ and $d_2$ polarization excitations in the photoluminescence spectra for 1.878eV excitation energy. Furthermore, a blue shift of approximately 3 meV in exciton energies relative to reflectance spectra was detected, which is likely attributable to hot photoluminescence processes and potentially accounts for the increased intensity observed for the higher-energy exciton[5,8,41]. To investigate the dependence of Raman mode intensities on excitation energy, the excitation energy was systematically tuned from approximately 1.54 eV to 1.66 eV in intervals of 0.01 eV. The intensity variation of the $A_g^{12}$ Raman mode as a function of excitation energy is presented for both $d_1$ and $d_2$ polarizations in Fig. 2(c) and 2(d), respectively, with corresponding emission energies indicated along the top axes for reference. For $d_1$ polarization excitation, two prominent peaks were observed at excitation energies of 1.54 eV and 1.58 eV. The 1.54 eV excitation corresponds to the pump resonance Raman scattering (pump-RRS) coinciding with the $X_1$ exciton. Since the $A_g^{12}$ phonon mode has an energy of ~40 meV, the Stokes RRS appears ~40 meV above the $X_1$ exciton, at an excitation energy of 1.58 eV. The observed Stokes-resonant and pump-resonant Raman scattering enhancements are approximately 200-fold and 140-fold greater than spontaneous Raman scattering, respectively. Similarly, for $d_2$ polarization excitation, two Lorentzian peaks were observed, with pump-RRS occurring at the $X_2$ exciton energy of 1.58 eV and Stokes-RRS observed at an excitation energy of 1.62 eV and the intensity is enhanced by ~100 and ~50 times with respect to spontaneous RS, respectively. The excitation energy dependent Raman intensity for the $A_g^7$, $A_g^{14}$ are shown in Fig. S2 of the

Supporting Information. The Raman scattering efficiency near the excitonic resonance can be described by the expression[39], $I_{RS} \approx \left|\frac{A}{|E_1+i\gamma_1-E_{exc}||E_2+i\gamma_2+E_v-E_{exc}|}\right|^2$ which has been used to fit the intensity variations shown in Fig. 2(c) and 2(d). Here, $E_1$ and $E_2$ represent energy of the intermediate states involved in the pump and Stokes resonance processes, respectively, while $\gamma_1$ and $\gamma_2$ are their associated damping constants. The numerator $A$ represents the Raman scattering matrix element, as detailed in Note S1 of the Supporting Information. The form of denominator suggests that Raman scattering intensity in $I_{RS}$ exhibits a double Lorentzian-like profile, centred at $E_{exc} = E_1$ and $E_{exc} = E_2 + E_v$, corresponds to pump and stokes RRS condition as mentioned above. Our fitting reveals that for a fixed polarization excitation the damping constant $\gamma_1$, $\gamma_2$ differ, even though the intermediate state energies coincide with the excitonic energy, i.e., $E_1 = E_2 = E_x$. For d$_1$ polarization excitation, the extracted values are

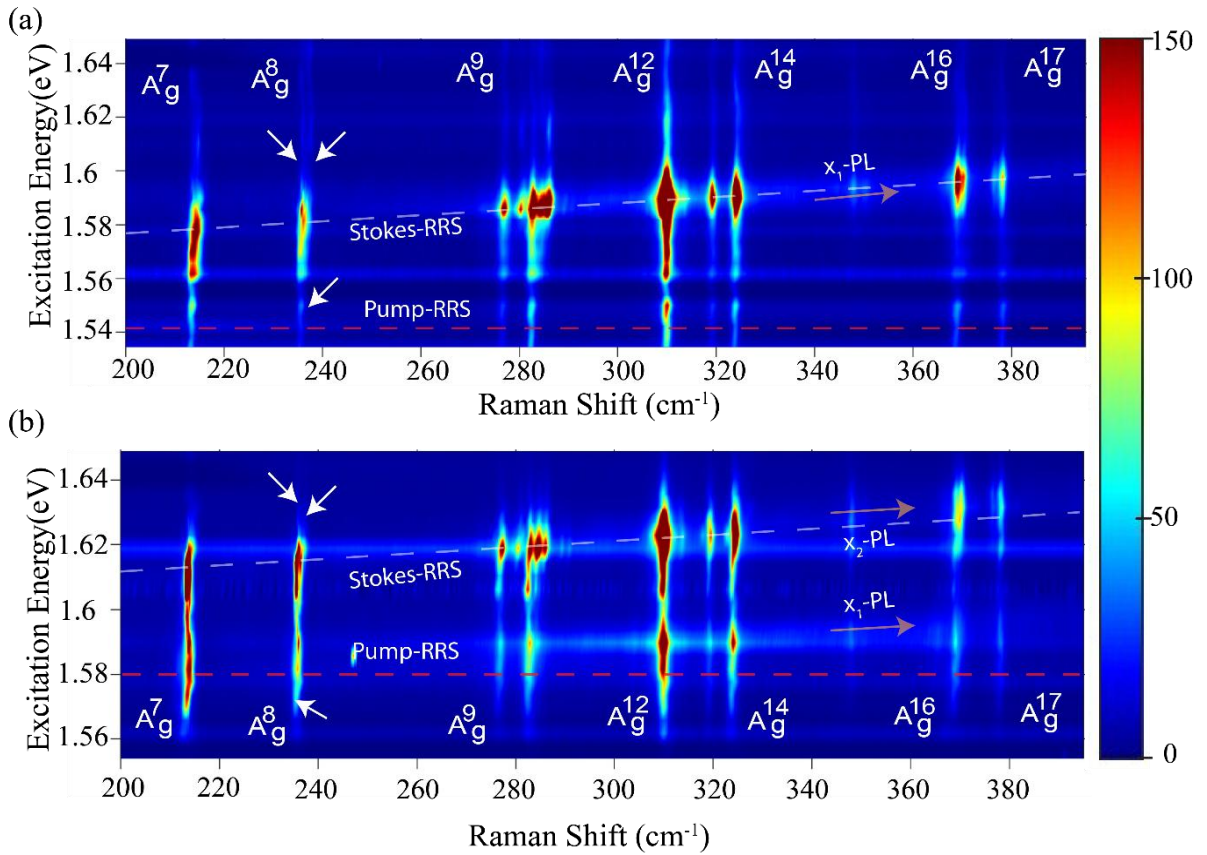

Fig.3 (a) Color map of emission spectra as a function of Raman shift for varying excitation energies under d$_1$ polarization excitation. (b) Same as (a), but for d$_2$ polarization excitation.

$\gamma_{x1}$ = 10.3 meV, $\gamma_{x2}$ = 8.57 meV, while d$_2$ polarized excitation, they are $\gamma_{x1}$ = 7.07 meV, $\gamma_{x2}$ = 5.56 meV. The distinct damping constants indicates that the effective intermediate state involved in the Raman process is not identical under pump- and Stokes-RRS conditions. This interpretation is further supported by the appearance of additional Raman modes observed specifically under the Stokes-RRS condition, as discussed in the following section. Fig. 3(a) and 3(b) present the complete emission spectra recorded under varying excitation energies for d$_1$ and d$_2$-polarized excitation configurations, respectively. We observed that, in the pump-RRS regime, a simultaneous enhancement of all phonon modes occurs at a fixed excitation energy, regardless of their individual vibrational energies. This collective enhancement appears as a horizontal region in the spectra, indicated by the red dotted line as a guide to the eye. In contrast, the Stokes-RRS regime exhibits phonon energy-dependent enhancement. Here, each Raman mode being enhanced at a different excitation energy depending on its vibrational energy. This behavior appears as an inclined region in the spectra, delineated by sloped dotted white lines. Furthermore, we observe the emergence of new Raman lines, under excitation energies above 1.585 eV and 1.62 eV for d$_1$- and d$_2$-polarized excitation, respectively. It has been theorized that such additional Raman peaks can appear when the intermediate state involved in the scattering process is not a Bloch state[42,43], such as when the process involves electrons of a bound excitonic state. Notably, our observation of two distinct intermediate state in pump and stokes RRS condition, are further validate by phonon assisted excitons photoluminescence. Under d$_1$ polarized excitation at 1.585 eV, a broad PL peak appears near the A1$_{2g}$ mode (Fig. 3a), identified as X$_1$ exciton PL emission. As the excitation energy increases, the Raman peaks shift according to energy conservation, while the PL features remain fixed in absolute energy. Consequently, when plotted on a Raman shift axis, PL signals appear to shift with a slope[44] (as illustrated by the arrow in Fig. 3a), while Raman modes remain at constant positions. Similarly, for d$_2$ polarization exctation , two broad peak are observed around 1.59 eV and 1.62 eV corrspons to $X_1$ and $X_2$ excitons photoluminescence (as indicated by the arrows in Fig. 3b). To gain a comprehensive understanding of photoluminescence (PL) as a function of excitation energy, the PL emission across the entire spectrum was deconvoluted, and the integrated PL intensity was plotted against excitation energy. Notably, under $d_1$-polarized excitation, the $X_1$-exciton PL begins to emerge at an excitation energy of approximately 1.581 eV around 40 meV above the exciton resonance energy and vanishes beyond ~1.593 eV as shown in Fig. 4(a). A comparable behaviour is exhibited by the $X_2$-exciton PL, which emerges at an excitation energy of approximately 1.61 eV (~40 meV above the $X_2$ exciton resonance) and is quenched near 1.635 eV, as depicted in Figure 4(b). The deconvoluted PL spectra, overlaid with Raman lines, are shown in Fig. S3 and Fig. S4 of the Supporting Information for d1 and d2 polarization excitations, respectively. The excitation energy dependent emergence and quenching of PL can be attributed to phonon-assisted exciton photoluminescence, as schematically illustrated in Figure 4(c). As $ReS_2$ is an indirect bandgap[8,38] semiconductor, phonon participation is essential for both exciton formation and radiative recombination. Notably, the phonon energy range corresponding to the full Raman spectrum lies between 20 and 50 meV. This energy window, located above the exciton resonance, is referred to as the $R_1$ region in the exciton dispersion. When the excitation energy

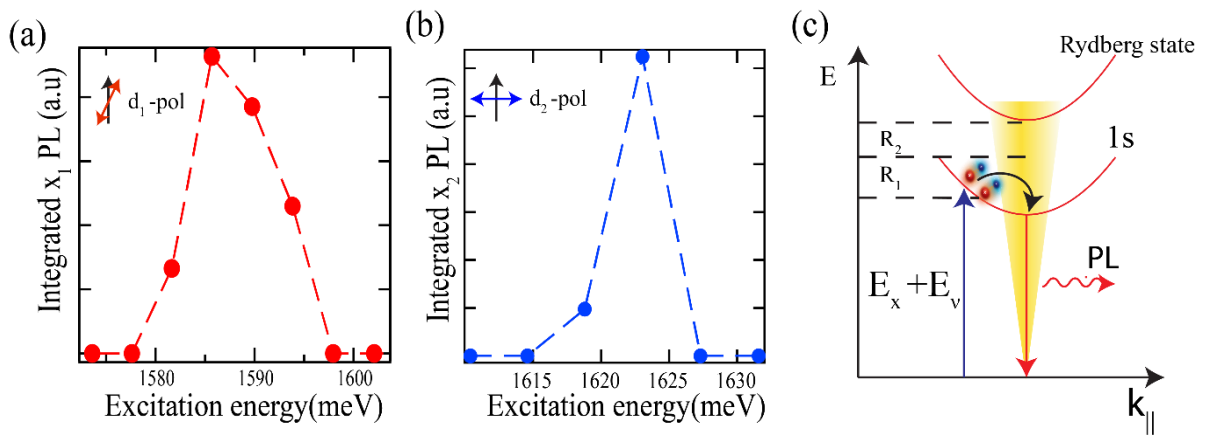

Fig.4 (a) Variation of integrated PL intensity of $X_1$ exciton under $d_1$ polarization excitation (b) variation of integrated PL intensity for the $X_2$ exciton under $d_2$ polarization excitation. (a) Schematic of exciton dispersion indicating the phonon energy range above the exciton resonance ($R_1$ region) and the energy gap between excitonic states ($R_2$ region).

falls within this region, excitons can be efficiently excited and recombines via phonon-assisted processes. As the excitation energy increases further, exceeding ~50 meV above the exciton resonance, a reduction in PL intensity is observed. This quenching may result from the reduced availability of phonons to facilitate exciton formation at higher excitation energies. Additionally, as the excitation energy approaches the Rydberg exciton regime, it likely enters the energy gap between the 1s and higher-order excitonic states[7], contributing further to the quenching effect, as indicated in the $R_2$ region. Furthermore, the PL linewidth is found to narrow significantly near resonance, measuring approximately 4-5 meV, in contrast to a broader linewidth of ~10 meV under far-from-resonance excitation. This linewidth narrowing is consistent with the phonon-assisted photoluminescence mechanism[45].

In conclusion, our study reveals that the presence of excitons has two major effect on Raman scattering in anisotropic $ReS_2$. First, Raman scattering has been enhance up to ~200 times compared to conventional spontaneous Raman processes. Secound, new Raman lines has been emerge under Stokes-resonant conditions. A detailed ab initio calculation of phonon dispersion is required for a quantitative understanding of these newly observed lines. Adittionally, the observation of phonon-assisted photoluminescence further supports the indirect bandgap nature of $ReS_2$. Overall, our findings establish resonance Raman scattering as a powerful tool for probing low-symmetry materials and guiding the development of Raman-based optoelectronic applications.

Methods:

Few-layer $ReS_2$ flakes were mechanically exfoliated from commercially available bulk crystals and dry-transferred onto a 340 nm $SiO_2$/Si substrate. A tunable Ti:sapphire laser (Mira 900) was employed for sample excitation. Optical measurements were conducted using a closed-cycle optical microscopy cryostat (Montana Instruments) with a variable temperature range of 3.2–295 K. All optical measurements were performed using a home-built microscope. A

Princeton Instruments spectrometer (SP2750) coupled with a liquid-nitrogen-cooled detector (PyLoN:400BR-eXcelon) was used to record the photoluminescence (PL) signal.

Acknowledgment:

This work has been supported by funding from the Science and Engineering Research Board (CRG/2018/002845); Ministry of Education (MoE/STARS-1/647); Council of Scientific and Industrial Research, India (09/081(1352)/2019-EMR-I) and Indian Institute of Technology Kharagpur.

# Supporting Information for
# Giant Resonance Raman Scattering via Anisotropic Excitons in ReS$_2$

Pritam Das, Devarshi Chakrabarty, Neha Gill, Sajal Dhara*

Department of Physics, Indian Institute of Technology, Kharagpur, India

**List of Contents:**

Figs. S1-S4

Note S1

References

**Note S1. Resonance Raman Scattering Near Exciton Resonance:**

The Raman scattering efficiency near excitonic resonances can be described by the following expression,

$$I_{RS} \approx \left(\frac{2\pi}{\hbar}\right) \left| \frac{\langle 0|H_{e-R}(\omega_S)|a\rangle\langle a|H_{e-ph}|a\rangle\langle a|H_{e-R}(\omega_p)|0\rangle}{|E_a - E_{pump} + i\gamma_1||E_a - E_{pump} + E_{phonon} + i\gamma_2|} + C \right|^2$$

where $H_{e-R}$ and $H_{e-ph}$ represent the electron–radiation and electron–phonon interaction Hamiltonians, respectively. The initial electronic state is assumed to be the semiconductor ground state $|0\rangle$, without electron–hole pair excitations. Upon photon excitation, the system transitions to an intermediate state $|a\rangle$ with energy $E_a$. Here, $E_{pump}$ and $E_{phonon}$ denote the pump photon and phonon energies, respectively, and C represents contributions from non-resonant scattering. If the intermediate state $|a\rangle$ corresponds to a Bloch state, it possesses a well-defined crystal momentum k. Under the dipole approximation, this restricts the light–matter interaction to involve only zone-centre phonons with momentum $q = 0$, ensuring that the phonon line shape remains unchanged even near resonance. In contrast, if $|a\rangle$ is not a Bloch state, such as in the case of a bound exciton, momentum is no longer a well-defined quantum number. As a result, the momentum conservation rule is relaxed[1–3], allowing phonons with finite momentum to participate in the scattering process, which can lead to modifications in the phonon spectrum, including the appearance of additional Raman modes and linewidth broadening.

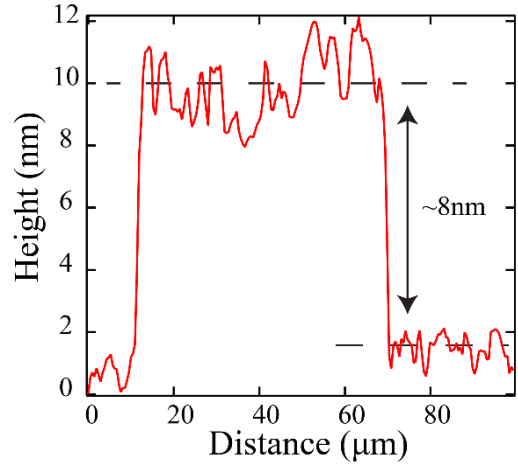

**Fig. S1.** Thickness profile of ReS$_2$ measured in Atomic Force Microscope along the the yellow line in optical microscope image as shown in fig. 1(a)

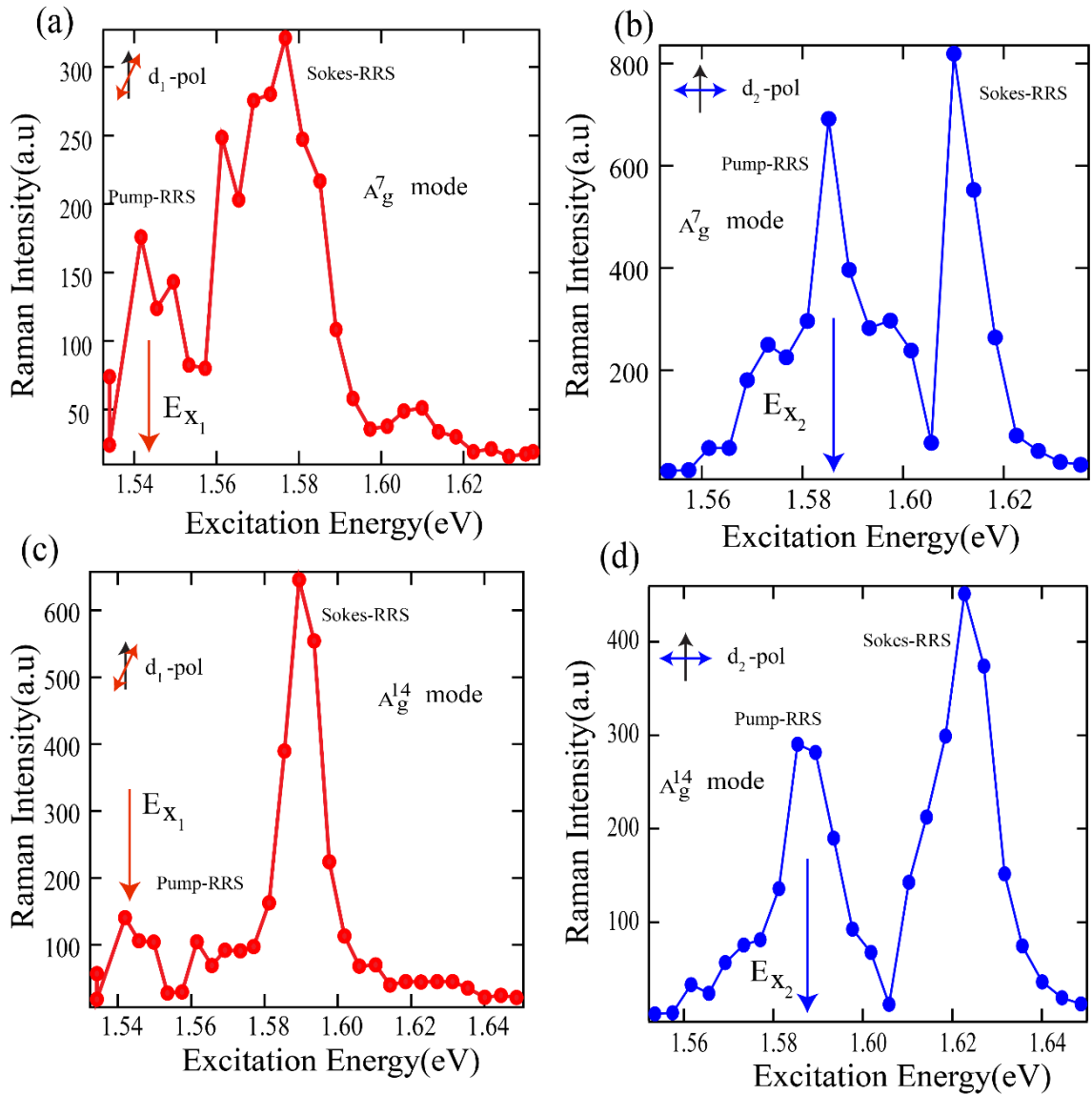

**Fig. S2.** (a) Variation of the integrated Raman intensity of the A7g mode as a function of excitation energy under d1 polarization. (b) Same as (a), but for d$_2$ polarization. (c) Variation of the integrated Raman intensity of the A14g mode with excitation energy for d$_1$ polarization. (d) Same as (c), but for d$_2$ polarization.

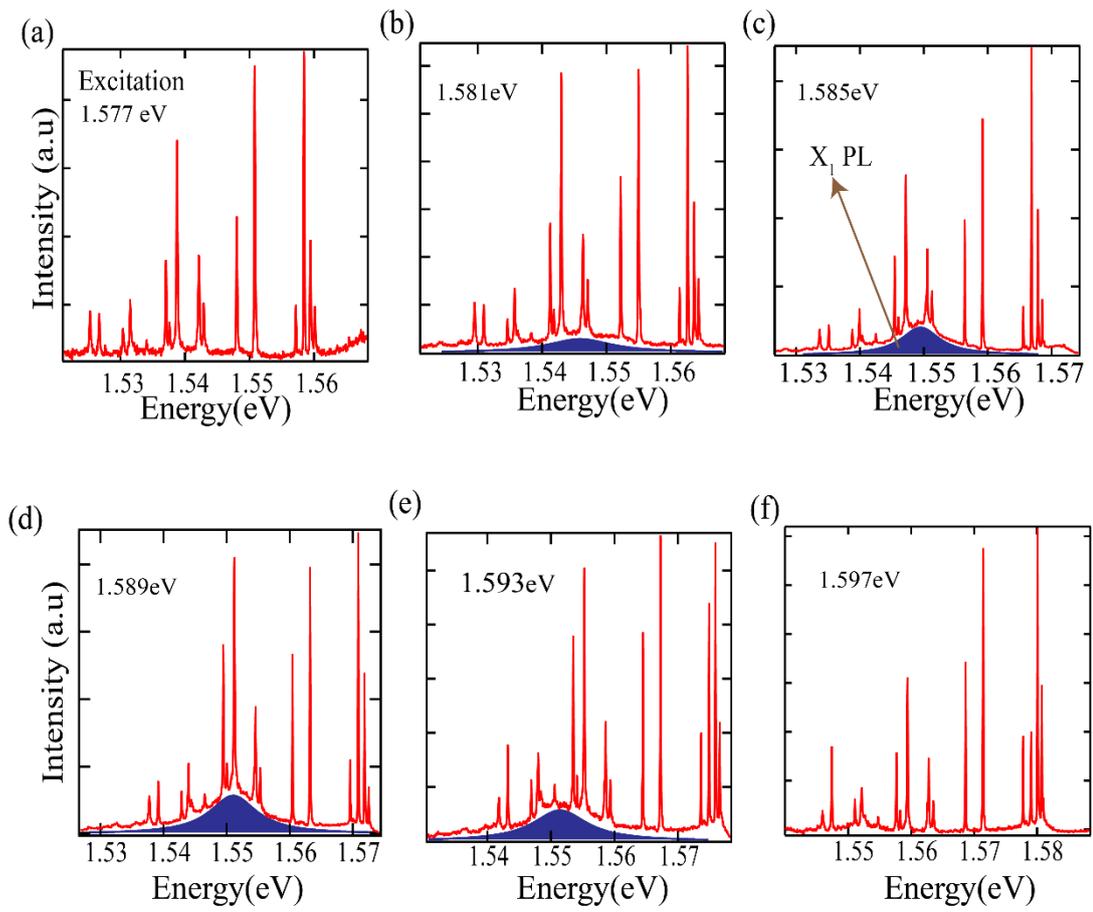

**Fig. S3.** (a–f) Emission spectra measured under $d_1$ polarization excitation at excitation energies of 1.577, 1.581, 1.585, 1.589, 1.593, and 1.597 eV. The deconvoluted photoluminescence (PL) spectra are overlaid on the emission spectra for the excitation energies from 1.581 to 1.593 eV, as shown in panels (b–e).

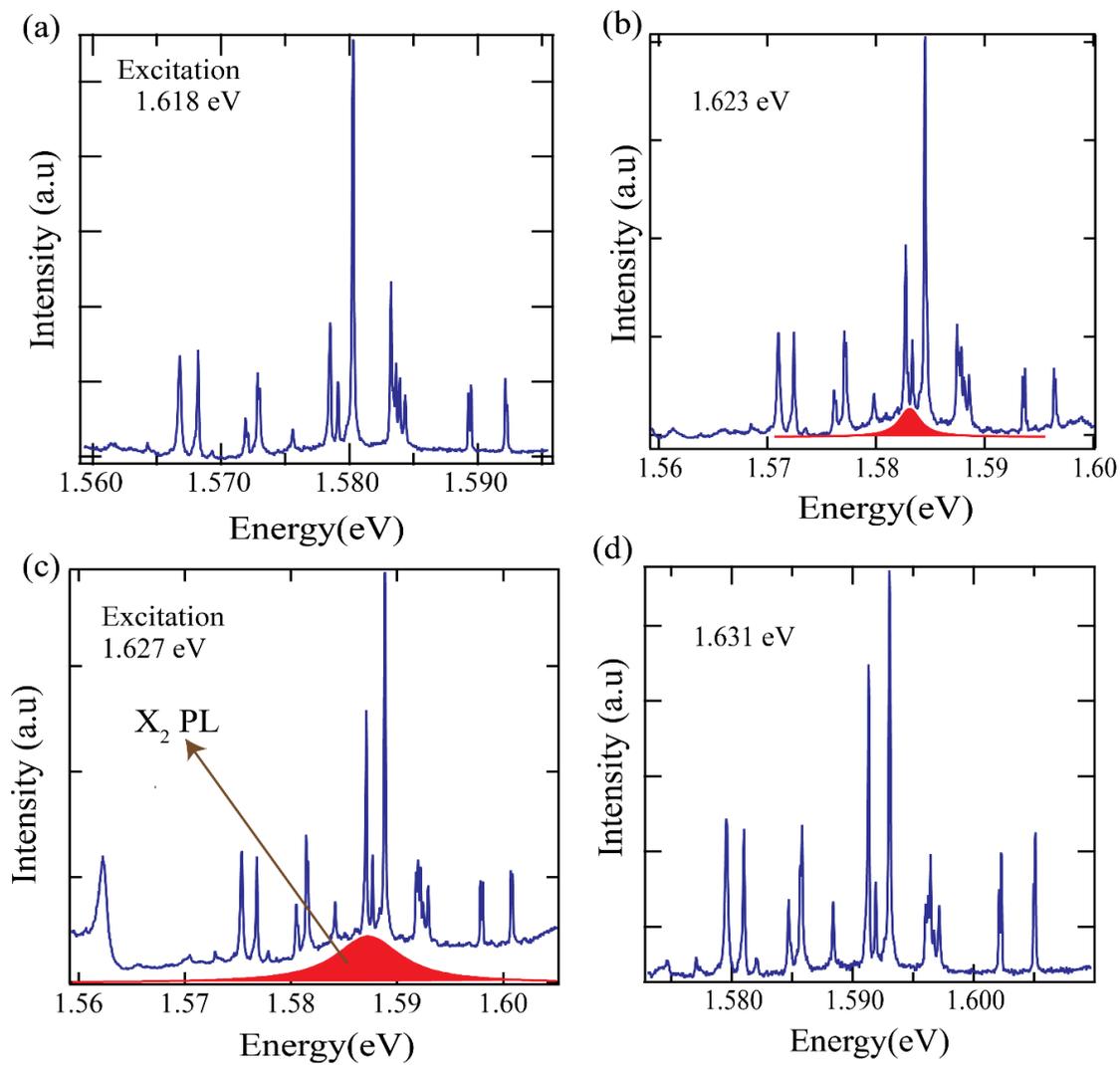

**Fig. S4.** (a–d) Emission spectra measured under $d_2$ polarization excitation at excitation energies of 1.618, 1.623, 1.627, and 1.631 eV. The deconvoluted photoluminescence (PL) spectra are overlaid on the emission spectra for the excitation energies 1.623 and 1.627 eV, as shown in panels (b–c).